# Regions In a Linked Dataset For Change Detection

Anuj Singh

GeoPhy, Phoenixstraat 28c, 2611 AL, The Netherlands

singha2@tcd.ie

**Abstract.** Linked Datasets (LDs) are constantly evolving and the applications using a Linked Dataset (LD) may face several issues such as outdated data or broken interlinks due to evolution of the dataset. To overcome these issues, the detection of changes in LDs during their evolution has proven crucial. As LDs evolve frequently, the change detection during the evolution should also be done at frequent intervals. However, due to limitation of available computational resources such as capacity to fetch data from LD and time to detect changes, the frequent change detection may not be possible with existing change detection techniques. This research proposes to explore the notion of prioritization of regions (subsets) in LDs for change detection with the aim of achieving optimal accuracy and efficient use of available computational resources. This will facilitate the detection of changes in an evolving LD at frequent intervals and will allow the applications to update their data closest to real-time data.

## 1    Problem Statement

*How can we frequently monitor the changes in an evolving Linked Dataset to maintain interlinks and synchronize replicas of the dataset, given limits upon computation resources like capacity to fetch data from Linked Dataset and time frame to identify changes?*

Many Linked Datasets (LDs) are highly dynamic in nature [1]. For an application consuming a Linked Dataset (LD), the dynamic nature of the dataset may result in several issues such as broken interlinks or outdated data [1].

To maintain interlinks or to synchronize local copies of an LD during its evolution, there is a need to identify the changed resources and changed triples in the dataset [1]. Different types of change information support different type of use cases: **Use case of structural interlink maintenance**[1] requires information on the resources in the LD that changed their IRIs or were deleted; **Use case of semantic interlink**

---

[1] An interlink is structurally broken, if either source or target is not accessible at its IRI.

**maintenance**[2] requires information on the resources in the LD that changed their representation; **Use case of synchronizing local copies** requires information of deleted and added triples. For identifying the aforementioned changes in an LD, the state-of-art approaches either periodically detect changes in the online version of the datasets [1] or compare the resources or triples present in dumps of two consecutive versions of the datasets [2].

Approaches [1] that periodically detect changes in the online version of an LD, usually store the initial state of LD ("initial dataset"), and periodically access all the resources/ triples in the online version for comparison with the resources/ triples in the initial dataset. While, the dump-oriented change detection approaches [2] access dumps of both versions to compare the resources/ triples of one version with another.

As LDs are highly dynamic in nature, the change detection needs to be done at frequent periods for maintaining interlinks and synchronizing replicas. However, due to limits upon computational resources, the change detection approaches may not be able to detect changes in LDs at frequent intervals [3]. As a result, the applications that consume LDs to provide services potentially will be doing so on top of outdated data or may no longer dereference the available data due to broken interlinks that might be created in the evolution of the datasets.

Research in [3], [4], [5] have identified the following change behavior in LDs during their evolution: the majority ~62% of Linked Data in LOD cloud is static (no change); some LDs are less likely to change and some LDs have a higher change rate; and resources in LDs representing some concepts have higher change rate than resources of other concepts.

Inspired by the identified change behavior of LDs, this research proposes using the notion of *regions* when frequent change detection is necessary in an evolving LD. This paper defines *"region" as a subset of an LD, which may constitute several concepts with their resources in which the same type of changes have occurred with similar frequency.* As described earlier, there are multiple types of change in LDs and different types of change support different types of use case. Hence, it should be interesting to see if regions can be identified in an LD using the frequency and type of changes. By doing so we should be able to *prioritize* regions according to different use cases, such that changes can be detected at shorter/ longer time-intervals in a specific region based on its priority. This in turn should allow better use of limits upon computational resources and achieve optimal accuracy.

To realise the new region-based approach for change detection, several complementary approaches/ methods are needed, which include: a change detection and classification mechanism to identify different type of changes in an LD; a method to identify the regions in an LD based on the frequency and type of changes; and a method to assign priority to each identified region for change detection.

---

[2] An interlink is semantically broken when the meaning of the representation of target differs from the intended meaning of the source.

## 2 Relevancy

Many applications provide services on top of large LDs such as DBpedia and LinkedGeoData [6]. For this, on one hand, the service providers like LocationLoder[3] link resources in their dataset with the resources in other LDs such as DBpedia. On the other hand, some service providers like sameAs.org[4] replicate the LDs to increase the flexibility of information sharing and integration infrastructure. These applications can be impacted by evolution of the interlinked or replicated LDs, as the evolution of an LD may cause both broken interlinks as well as outdated replicas.

For maintaining broken interlinks or outdated replica of an LD, the state-of-art suggests identifying different types of changes during the evolution of the dataset, which are as follows: deleted resources; resources that changed their IRIs; resources that changed their representation; added triples; deleted triples.

**What is the research problem here?** The necessity for approaches that *efficiently* identify changes in an evolving LD and support use cases like automatic repair of broken interlinks and synchronization of replicas, is a research problem that has also been highlighted by Popitsch et al. [1] and Umbrich et al. [7].

**What would happen if the proposed approach succeeds?** The proposed approach will make it easier to detect changes in LDs efficiently. Hence, applications providing services on top of LDs will be more effective and easier to build and maintain.

## 3 Related Work

This section presents the research related to the analysis of LOD cloud to explore the characteristics of its constituent Linked Datasets (LDs).

In order to analyze the dynamics of Linked Data on Web, Umbrich et al. [8] extracted the LD documents using the MultiCrawler framework. This framework extracted the data from 7 hop neighborhood of Tim Berners-Lee's FOAF file[5] for 24 weeks resulting in 550,000 RDF/ XML documents. While, Kafer et al. [5] extracted LD documents by a crawler using: 220 URIs available on the DataHub site under the "LOD cloud" group; and top 220 URIs extracted from BTC 2011 dataset. Kafer et al. then performed a 2-hop breadth first crawl using the 440 URIs as the seed-list, which resulted in 95,737 dereferenceable URIs, documents on which comprise the data of 652 LDs. Kafer et al. then monitored these documents for 29 weeks. The number of unique documents that appeared in at least one snapshot was 86,696. Both Umbrich et al. and Käfer et al. analyzed their respective documents and identified that ~62% of the total documents did not change at all. Umbrich et al. also mentioned that half of the documents that changed had a change frequency of more than 3 months. Käfer et al. added to this by mentioning that 23.2% of the total monitored documents were changing infrequently. Having the data analyzed of more than 600 LDs, the findings of Kafer et al. also suggest that the data of 51.9% LDs did not change in 29 weeks.

---

[3] http://apps.dri.ie/locationLODer/locationLODer

[4] http://sameas.org/

[5] https://www.w3.org/People/Berners-Lee/card

In 2014, Dividino et al. [9] monitored 13 LDs in LOD cloud for 19 months and analyzed their change rate. The results showed that out of 13 LDs used for analysis, 5 LDs were highly dynamic, which implies their high change rate during the entire monitored time period. Another 5 LDs were identified as less dynamic as these LDs had lower change rate during the entire monitored time period. The rest of the LDs were also identified as highly dynamic but the change rate of these LDs was not high during the entire monitored period. These LDs were identified as highly dynamic as at the latest points in the monitored time period these LDs changed rapidly. Based on these findings, and the findings of Käfer et al. [5], in 2015, Dividino et al. [3] proposed a scheduling strategy for replica synchronization based on the change rate of LDs. Inspired by the research of Dividino et al., in 2017, Nishioka et al. [10] proposed a novel scheduling strategy for crawling data from LOD cloud to synchronize the local copy of LOD data used by applications. The proposed scheduling strategy is built on top of the study of the triples in DyLDO dataset and Wikidata. Nishioka et al. analyzed 173 weekly snapshots of DyLDO dataset and 25 monthly snapshots of Wikidata to identify the ephemerality and stability of the triples in these datasets.

The approaches discussed so far have analyzed multiple LDs/ data in LOD cloud. However, none of these approaches analyzed a particular LD in isolation. In 2011, Wang et al. [4], studied individual LDs for concept drift. Wang et al. analyzed all the resources of the concepts of DBpedia versions 3.2, 3.3, 3.4 and 3.5. The approach compared the resources in a concept with the resources of the same concept in different versions. This led to the identification that some of the concepts of DBpedia are very unstable (such as OfficeHolder, Politician, City, College, ChemicalCompound) while few concepts are very stable (Planet, Road, Infrastructure, Cyclist, LunarCrater) in nature. Stability has been calculated by the change in the resources (added or deleted) of a particular concept in various versions. The findings of Wang et al. suggest that different concepts in Linked Dataset could evolve differently.

## 4 Reflection

**Rationale for the success of a new region-based approach for change detection:** As discussed in the related work, the state-of-art approaches have identified following patterns during the evolution of LDs: 1) Majority (~62%) of the data in LOD is stable in terms of changes. 2) The Data that does change also changes in pattern i.e. partly frequently, while the other part changes infrequently. 3) Using the information of changed resources, the existing studies [4] have also identified that different concepts can have different change rate. Based on this, it can be argued that different concepts evolve differently. Hence, different regions can also be identified in an evolving LD.

Many successful scheduling strategies [3], [10] have been designed and evaluated based on pattern 1 and 2 for crawling data from LOD to synchronize the local copies of the LDs. However, to the best of my knowledge, there is no published study that identifies regions (concept(s)) within an LD based on frequency and type of changes, and uses the identified regions for designing the scheduling strategies for monitoring the changes in an evolving LD.

The identification of pattern 3 by existing studies [4] suggest that scheduling strategies for monitoring changes in an evolving LD could successfully be built on top of different regions in the dataset.

## 5  Research Question

*Using the regions identified based on frequency and type of changes in a Linked Dataset, to what extent can we prioritize various regions of a Linked Dataset (LD) for change detection in order to achieve optimal accuracy?*

Where *accuracy* denotes the F-measure of a change detection mechanism in identifying the changes in an LD, and *optimal accuracy* denotes the best possible F-measure that can be achieved by applying the change detection mechanism only on the dynamic regions of the LD.

To answer the research question, the following research objectives have been derived:

**Research Objective 1:** Design and evaluate a change detection and classification method (CDCM) for LDs to improve the accuracy of existing CDCMs.

**Research Objective 2:** Design and evaluate an aggregation method (AM) to identify regions based on frequency and type of changes in an LD during its evolution.

To define regions formally, it is important to have a formal definition of LD as regions are the subsets of an LD. The following definition of LD is in terms of concepts and resources (instances), this definition is based on the notion of the linked RDF dataset provided by Alexander et al. [13].

***Definition 1*** **(LD):** *Consider a set of concepts C, and set of resources (instances) R, using these notations, an LD can be formally characterized by the following, LD = {r | r rdf:type c, r ∈ R, c ∈ C}.*

That is, an LD is a set of resources (instances) classified under some concepts that are published and maintained by a single provider.

*Definition 2* (**Regions**): *Consider a set of concepts C, set of resources R, and a set of types of change Δ. The frequency of single type of change in the resources of a concept can be determined by function f(δ), where δ ∈ Δ, and definition 1 describes the relation between resources and concepts in an LD. Using f(δ), set of concepts C' can be determined whose resources (instances) have experienced same type of changes that occurred with similar frequency, where C'⊆ C. Then, the region for δ type of change can be formally characterized by Reg(f(δ)) = {r| r rdf:type c,  r ∈ R, c ∈ C'}, where Reg denotes the region in which δ type of changes have occurred with frequency f(δ).*

**Research Objective 3:** Design and evaluate a scheduling method (SM) to prioritize the regions of an LD for change detection during the evolution of the dataset.

# 6 Approach

The main idea behind the approach is to provide a way to frequently monitor/ detect changes in an evolving LD using limited computational resources such as: limited capacity to fetch data from LD; short time frame to compute changes. The key innovation of this research is the identification of regions in an evolving LD by analyzing the change behavior (frequency and type of changes) of concepts that evolve differently during the evolution of the dataset. The identified regions can then be prioritized for change detection, such that changes can be detected at shorter/ longer time-intervals in a specific region based on its priority.

To fulfill research objective 1, the idea is to propose a method that detects and classifies the changes in an evolving LD when provided two versions (v1 and v2) of the dataset as input. For this, DELTA-LD, an approach that detects and classifies the changes between two versions of an LD has been proposed [14]. DELTA-LD, focuses on detecting and classifying the changes in the context of use cases where automatic repair of broken interlinks or synchronization of replica of an LD is required. There are other approaches [15], [16] that produce changes at more granular level. However, this granularity adds more human understandability to the detected changes, which in the considered use cases (maintenance of interlinks and replica) is not essential. DELTA-LD detects the following types of change: **(a)** newly added resources; **(b)** deleted resources; **(c)** resources that have had their representation changed; **(d)** resources that have had their IRIs changed; **(e)** resources that have had both their IRI and representation changed; **(f)** added or deleted triples. DELTA-LD has been evaluated by an experiment, details of which is mentioned in Experiment 1 (section 7.1).

To address research objective 2, the idea is to design a method that performs following three operations: **(a)** Aggregate the changes of consecutive sets of two versions of an LD ({v1, v2}, {v2, v3}, … {vn-1, vn}) till version 'n'. This will result in the information of different types of changed resources over a period of time along with the information of concepts to which the changed resources belong. **(b)** Apply an event distribution model on the aggregated changes to assign a probability to each concept for each type of change based on the frequency of that change in the concept. **(c)** Once the probabilities for each concept have been assigned, the method will then group the concepts with similar probabilities into a region. This method will be evaluated by an experiment. For details about the experiment, see Experiment 2 (section 7.2).

To address research objective 3, a method needs to be designed that prioritizes different regions in an evolving LD by taking the objective of change detection and dynamics (basically the probability attached to the constituent concepts) of regions as input. This method will be evaluated by an experiment. For details about the experiment, see Experiment 3 (section 7.3).

## 7   Evaluation Plan

To evaluate the accomplishment of the research objectives mentioned in section 5, the plan is to conduct following three experiments.

### 7.1 Experiment 1

**Purpose:** To determine the accuracy of proposed CDCM and compare it with existing change detection approaches [1], [11]. As, Popitsch et al. [1] and Pourzaferani et al. [11] use similar change metrics as the proposed CDCM, thus, led their selection.

**Hypothesis (H1)**: Given two versions of an LD as input to detect changes, the accuracy of proposed CDCM will outperform the existing change detection approaches [1], [11] in terms of F-measure.

**Input datasets:** Enriched DBpedia person snapshots 3.2 and 3.3 provided by Popitsch et al. [1]; and DBpedia person snapshots 3.6 and 3.7;

**Gold Standard:** The plan is to use the following gold standards (GS): GS provided by Popitsch et al. [1] for the changes between DBpedia snapshots 3.2 and 3.3; for the changes between snapshots 3.6 and 3.7, manually engineered GS will be used, creation of which includes DBpedia disambiguation dataset 3.7, DBpedia redirect dataset 3.7, and manual verification of the detected changes, which will not be found in the redirect and disambiguation dataset;

### 7.2 Experiment 2

**Purpose:** Evaluate the accuracy of the designed AM in identifying regions in an LD based on the frequency and type of changes.
**Hypothesis (H2):** In cases, where there are multiple types of changes in the resources of a concept, out of all the probabilities attached to these concepts for each type of change, the probability for *added resources* will be the maximum in most cases.
**Input datasets:** DBpedia versions 3.2, 3.3, 3.4, 3.5, 3.6, 3.7, and 3.8; Wikidata weekly snapshot for 3 months.
**Gold Standard:** Manual verification; findings of Wang et al. [4]; BEAR-B dataset [12], which contains the change set of DBpedia Live for August to October 2015.

### 7.3 Experiment 3

**Purpose:** To evaluate the efficiency of the proposed SM for change detection.
**Hypothesis (H3):** Given a certain capacity to fetch data from LD and a time frame to perform change detection, the proposed SM will outperform the existing scheduling strategies (the GS to be used) in terms of optimal accuracy of the detected changes.
**Input:** Information of regions identified by AM; all previous versions of DBpedia. Weekly snapshots of Wikidata for last 3 months (from the date of experiment); Live versions of DBpedia and Wikidata.
**Gold Standard:** Existing scheduling strategies that are based on age, size, importance, and change ratio; DBpedia live version change set; Wikidata live version change set.

## 8  Preliminary Results

At this stage, research objective 1 (see section 5) has been achieved. This section presents the results identified on the accomplishment of research objective 1.

DELTA-LD, a change detection and classification method has been proposed that detects resource level changes along with their changed triples between two versions of an LD using the following classification of changes: *create* – a new resource has been added in LD; *remove* – an existing resource has been deleted from LD; *update* – representation of an existing resource in LD has been changed; *move* –IRI of an existing resource in LD has been changed; *renew* – both IRI and representation of an existing resource in LD has been changed [14].

Experiment 1 (see section 7.1) was conducted to evaluate the accuracy of DELTA-LD. The **input datasets** were used in two sets, and each set contains two snapshots of DBpedia, following are the details: *set 1* – Enriched DBpedia person snapshots 3.2 and 3.3 (20,284 and 29,498 resources) provided by Popitsch et al. [1]; *set 2* - DBpedia person snapshots 3.6 and 3.7 (296,595 and 790,703 resources). Details of **Gold standard (GS)** and **hypothesis** are mentioned in section 7.1. **Methodology:** The proposed DELTA-LD approach has been applied to snapshots of set 1 and 2 to detect and classify the changes between the snapshots. **Discussion:** Table 1 describes the detected and classified changes for snapshots of set 1 and 2, which are also available online[6]. In comparison to the existing classification of changes by Popitsch et al. [1], the classification of changes by DELTA-LD allowed to identify 46 and 1529 additional resources that changed their representation for set 1 and 2 respectively.

**Table 1.** Detected and classified changes for set 1 and 2

| Input Set | Create | Remove | Update | Move | Renew |
|---|---|---|---|---|---|
| 1 | 3819 | 239 | 4161 | 124 | 46 |
| 2 | 499590 | 5482 | 50380 | 2723 | 1529 |

To determine the accuracy of DELTA-LD, only move and renew types of changes have been compared with GSs, as the state-of-art emphasizes on determining the accuracy for detecting the resources that changed their IRIs. Since GSs do not cater move and renew types of changes separately, the detected move and renew types of changes have been merged as move type of changes. Move and renew types of change will now be referred to as move type of change in this section. Table 6 describes the accuracy of DELTA-LD in detecting move type of changes.

**Table 2.** Accuracy of DELTA-LD in detecting move type of changes

| Input Set | Move in GS | Move by DELTA-LD | Precision | Recall | F-Measure |
|---|---|---|---|---|---|
| 1 | 180 | 170 | 1 | 0.9444 | 0.9714 |
| 2 | 4271 | 4252 | 0.9597 | 0.9555 | 0.9576 |

Popitsch et al. [1] and Pourzaferani et al. [11] evaluated their approaches using snapshots of set 1. Popitsch et al. detect changes in an LD by periodically monitoring the online version of the dataset. So, in their conducted experiment, during the entire monitoring period, the maximum recorded precision, recall, and F-measure were 1.0, 0.91, and 0.95 respectively. The approach of Pourzaferani et al. compares two versions of an LD. In their experiment, the recorded precision, recall, and F-measure were 0.87, 0.99, and 0.93. By comparing the accuracy of DELTA-LD with the accuracy of Popitsch et al. and Pourzaferani et al., it has been identified that DELTA-LD outperforms the former by ~2% and the latter by ~4% in terms of F-measure, hence, the results support the formulated hypothesis (section 7.1).

---

[6] https://github.com/anujsinghdm/DELTA-LD/tree/master/Results

The accuracy of DELTA-LD for set 2 was not compared to any other approach, as to the best of my knowledge, only Pourzaferani et al. [11] have published their results of detected move type of changes on snapshots of set 2. However, the GS used by them has different count than the GS used to identify the accuracy of DELTA-LD. They neither published the details of creating their GS nor they published their GS as yet. This prevents the comparison of the results of set 2 with other approaches at this time.

**Acknowledgements.** I thank my supervisors Prof. Declan O'Sullivan and Dr. Rob Brennan at ADAPT Centre for Digital Content Technology.